\documentclass[prb,twocolumn,showpacs]{revtex4}
\usepackage{graphics}

\newcommand{\Journal}[4]{#1 \textbf{#2}, #3 (#4)}

\newcommand{\pzt}{$\mbox{PbZr}_{1-x}\mbox{Ti}_x\mbox{O}_3$}
\newcommand{\pzn}{$\mbox{Pb(Zn}_{1/3}\mbox{Nb}_{2/3})_{1-x}\mbox{Ti}_x\mbox{O}_3$}
\newcommand{\be}{\begin{equation}}
\newcommand{\ee}{\end{equation}}

\begin{document}

\title{Phenomenological theory of phase transitions in highly piezoelectric
perovskites}
\author{I. A. Sergienko}
\email[E-mail:]{iserg@uic.rsu.ru}
\author{Yu. M. Gufan}
\affiliation{Institute of Physics, Rostov State University,
Rostov-on-Don, 344090, Russia}
\author{S. Urazhdin}
\affiliation{Department of Physics and Astronomy, Michigan State
University, East Lansing, MI 48824}

\begin{abstract}
Recently discovered fine structure of the morphotropic phase
boundaries in highly piezoelectric mixture compounds PZT, PMN-PT,
and PZN-PT demonstrates the importance of highly non-linear
interactions in these systems. We show that an adequate
Landau-type description of the ferroelectric phase transitions in
these compounds is achieved by the use of a twelfth-order
expansion of the Landau potential in terms of the phenomenological
order parameter. Group-theoretical and catastrophe-theory methods
are used in constructing the appropriate Landau potential. A
complete phase diagram is calculated in phenomenological parameter
space. The theory describes both PZT and PZN-PT types of phase
diagrams, including the newly found monoclinic and orthorhombic
phases. Anomalously large piezoelectric coefficients are predicted
in the vicinity of the phase transition lines.
\end{abstract}


\maketitle

\section{Introduction}\label{sec1}
For many years, perovskite-type materials have been a subject of
extensive research in both experimental and theoretical physics.
On one hand, different representatives of the perovskite family
exhibit a host of physical phenomena, such as piezoelectricity,
ferroelectricity, and superconductivity; on the other hand,
perovskite structure is a relatively simple and, thus, attractive
object for theoretical studies.

Even though there is a long history of studies of perovskites,
they still present new surprises. Recent X-ray and neutron
diffraction studies on solid solutions \pzt(PZT),
Pb(Mg$_{1/3}$Nb$_{2/3}$)$_{1-x}$Ti$_x$O$_3$ (PMN-PT), and \pzn
(PZN-PT)\cite{noh1,noh2,noh3,noh4,ye} have revealed new phases in
the vicinity of the morphotropic phase boundary\cite{jaffe} on the
$T-x$ phase diagram of the solutions. In a narrow Ti concentration
range $(x=46-52\%)$, the low temperature structure of PZT was
found to be monoclinic M$_A$ (crystallographic symmetry $Cm$) with
polarization vector ${\mathbf P}$ directed along $[uuv], u<v$
pseudo-cubic direction\cite{noh3}. A similar M$_A$ structure has
also been recently seen in PMN-PT below room temperature for
$x=35\%$~\cite{ye}. Another fine structure of the morphotropic
phase boundary has been found in PZN-PT. For $x=9-11\%$, the
low-temperature structure is orthorhombic (O, $Amm2$) with
$\mathbf{P}\parallel[101]$~\cite{noh4}. Rhombohedral (R, $R3m$)
unpoled crystal of PZN-PT ($x=8\%$) was also found to exhibit
irreversible monoclinic M$_C$ ($Pm$, $\mathbf{P}\parallel[0uv]$)
distortion when an electric field above a certain critical value
is applied along the [001] pseudo-cubic direction~\cite{noh5}.

Early theoretical investigations of phase transitions in
perovskites were concentrated on BaTiO$_3$, which goes  through a
sequence of phases upon cooling: cubic(C, $Pm3m$), tetragonal(T,
$P4mm$), O, and R. Devonshire~\cite{devon} explains the behavior
of BaTiO$_3$ within the framework of a phenomenological
Landau-type expansion up to sixth order in terms of the
ferroelectric order parameter -- polarization $\mathbf{P}$. While
successfully describing the phase diagram of BaTiO$_3$, the
potential used in Ref.~\onlinecite{devon} lacks the high-order
terms necessary to describe the low symmetry phases. Using a
geometric argument based on the Curie principle, Zheludev and
Shuvalov~\cite{shuv} classified possible positions of $\mathbf{P}$
with respect to the cubic unit cell. Due to the purely symmetric
nature of this approach, it fails to distinguish between
monoclinic phases M$_A$ and M$_B$ (${\mathbf P}
\parallel[uuv], u>v$), because these phases have the same
crystallographic symmetry $Cm$. The group-theoretical relationship
between the geometric method and Landau approach was established
in Refs.~\onlinecite{guf, book}.

Using this approach, Gufan and Sakhnenko~\cite{sakh} found that on
a two-dimensional (e.g. $T-x$) phase diagram of perovskites there
can be a point $(T_0, x_0)$ where  five phases C, R, O, M$_C$, and
T co-exist. They calculated the phase diagram in the vicinity of
this five-phase point. Only small solutions of the equations of
state that are close to the five-phase point were considered,
justifying expansion in powers of small parameters $(T-T_0),
(x-x_0)$. However, the results of this work do not apply to the
recently discovered phases of the mixture compounds of Pb-based
complex oxides since, in this case, the R--O--T and R--M$_A$--T
triple points are separate from the C--R--T triple point.
Therefore, theoretical consideration cannot be limited to small
solutions of the equations of state, especially for the lowest
symmetry phases.

\emph{Ab initio}~\cite{bell}, as well as
phenomenological~\cite{vand}, calculations, have been used to
account for the presence of monoclinic phases on the $T-x$ phase
diagrams of ferroelectric perovskites. Vanderbilt and
Cohen~\cite{vand} calculate the phase diagram in the space of
phenomenological parameters within the framework of
Landau-Devonshire theory. Their model is based on the eighth-order
expansion of the Landau potential in terms of the polarization
orientation~${\mathbf P}/|{\mathbf P}|$. Although monoclinic
phases appear in the phase diagram of the model, it does not
incorporate cubic and triclinic (Tri, $P1$) phases.

Having considered a number of successively more complicated
models, a natural question arises: What is the most general
phenomenological model of the phase diagram of a cubic system
induced by a ferroelectric order parameter? This question can be
answered by use of the concept of \emph{integrity rational basis
of invariants} (IRBI), introduced in Appendix A. IRBI can be
thought of as a basis in the space of polynomials (formed from the
order parameter components), which are invariant under the
transformations of the symmetry group of the system. By a
group-theoretical argument, if the IRBI contains polynomials of
maximal order $n$, then at least a $2n$-th-order phenomenological
model is necessary to describe all the possible phases induced by
the order parameter~\cite{proh}. This statement is true for the
irreducible representations of groups generated by reflections,
including the $Pm3m$ symmetry of the perovskite structure. We will
show that, in the case of perovskites, the Landau potential has to
be expanded up to the twelfth-order terms to describe the phase
diagram induced by the ferroelectric order parameter.

The results of the analysis of a simple twelfth-order model are
presented in this paper. In Sec.~\ref{sec2}, we present the
solutions resulting in a phase diagram containing all the phases
allowed by the symmetry of the ferroelectric order parameter. In
Sec.~\ref{sec3}, we fit the $T-x$ phase diagrams of PZT and
PZN-PT. In Sec.~\ref{sec4}, we discuss the piezoelectric
properties of the compounds in the vicinity of the newly found
phase boundaries. Appendices are intended to give a
group-theoretical and catastrophe-theory background for some
general statements we make in the text.

\section{Twelfth-order Landau-type model}
\label{sec2}

Three algebraically independent $Pm3m$-invariant polynomials can
be formed (see Appendix A) from the polarization vector components
$(P_x, P_y, P_z)$.

\begin{equation}
\label{irbi}
\begin{array}{l}
J_1=P_x^2+P_y^2+P_z^2, \\
J_2=P_x^2P_y^2+P_y^2P_z^2+P_x^2P_z^2,\\
J_3=P_x^2P_y^2P_z^2.
\end{array}
\end{equation}

The ferroelectric part of the Landau potential can then be
expressed in terms of algebraic combinations of $J_1, J_2, \text{
and } J_3$. Since polynomials of up to sixth order are present in
the IRBI~(\ref{irbi}), the Landau potential has to be expanded up
to the twelfth order terms:

\begin{widetext}
\begin{equation}
\label{full12}
\begin{array}{llll}
F&=&a_1J_1+a_2J_1^2+b_1J_2 & \text{-- 2nd and 4th order terms}\\
&& +a_3J_1^3+d_{12}J_1J_2+c_1J_3 & \text{-- 6th order terms}\\
&& +a_4J_1^4+d_{112}J_1^2J_2+b_2J_2^2+d_{13}J_1J_3 & \text{-- 8th order terms}\\
&& +a_5J_1^5+d_{1112}J_1^3J_2+d_{122}J_1J_2^2+d_{113}J_1^2J_3+d_{23}J_2J_3 & \text{-- 10th order terms}\\
&& +a_6J_1^6+d_{11112}J_1^4J_2+d_{1122}J_1^2J_2^2
+b_3J_2^3+d_{1113}J_1^3J_3+d_{123}J_1J_2J_3+c_2J_3^2 &  \text{--
12th order terms}
\end{array}
\end{equation}
\end{widetext}

A complete investigation of extrema of the function $F({\bf P})$
in the multi-dimensional parameter space is a rather tedious
exercise. However, main features of the phase diagram can be
obtained from simplified models based on potential~(\ref{full12})
with some terms omitted. The truncated potential should satisfy at
least two requirements:  i) it has to be bounded from below, and
ii) it should be structurally stable in catastrophe theory
sense~\cite{arnold,proh}. The latter requirement provides that
small perturbations, which can arise from the omitted
in~(\ref{full12}) terms, do not drastically change the results
obtained in a simplified model. In Appendix~\ref{apb}, we briefly
describe how compliance with the second requirement can be
verified.

A twelfth-order model
\begin{equation}
\label{prim}
F_{init}=a_1J_1+b_1J_2+c_1J_3+a_2J_1^2+b_2J_2^2+c_2J_3^2
\end{equation}
meets both requirements when $a_2, b_2,$ and $c_2$ are positive
quantities, while $a_1, b_1,$ and $c_1$ are parameters driving
phase transitions, and can be of any sign. In spite of its
simplicity, model~(\ref{prim}) is in full agreement with the
results of the group-theoretical analysis of the ferroelectric
phase transitions in perovskites. It gives an account of all the
phases C, T, O, R, M$_A$, M$_B$, M$_C$, and Tri, and it correctly
describes the phase boundaries. For example, although the symmetry
groups of the phases M$_A$(M$_B$) and R obey a group-subgroup
relation $Cm\subset R3m$, phase transitions R--M$_A$ and R--M$_B$
cannot be of second order~\cite{book,vand}.

\begin{figure*}
\includegraphics{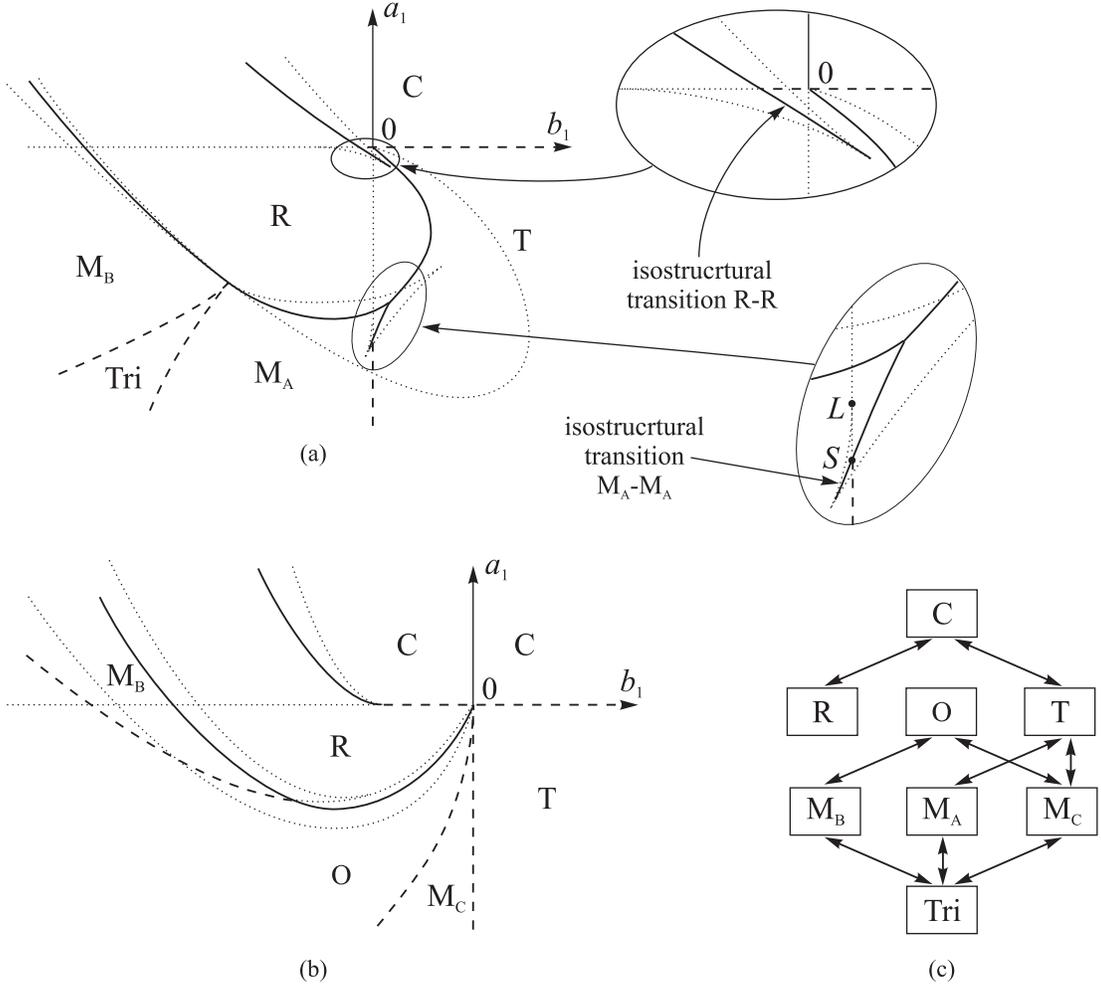}
\caption{\label{fig1}(a), (b) Phase diagram in $a_1b_1$-plane for
the potential $F_{init}$ for $c_1<0$(a) and $c_1>0$(b). Solid
lines -- first-order phase boundaries; dashed lines --
second-order phase boundaries; dotted lines -- stability
boundaries of phases. (c) Diagram of possible second-order phase
transitions between the phases.}
\end{figure*}

A complete phase diagram of the Landau potential~(\ref{prim}) can
be constructed in the space of phenomenological parameters
$(a_1,b_1,c_1)$. All the characteristic features can be seen from
the two-dimensional cross-sections in the $a_1b_1$-plane, as shown
in Fig.~\ref{fig1}.

Below we show how the phase diagram of Fig.~\ref{fig1} was
obtained, using the example of phase R. Minimizing the
potential~(\ref{prim}) with respect to $P_x, P_y, P_z$, and then
imposing the condition $P_x=P_y=P_z=P_s/\sqrt{3}$, we obtain the
value of spontaneous polarization $P_s$ as a solution of the
equation of state

\be \label{st}
a_1+\frac{2}{3}(3a_2+b_1)P_s^2+\frac{c_1}{9}P_s^4+\frac{4b_2}{9}P_s^6+\frac{2c_2}{243}P_s^{10}=0
\ee that obeys the stability conditions
\begin{eqnarray}
\label{cond1}
(3a_2+b_1)+\frac{c_1}{3}P_s^2+2b_2P_s^4+\frac{5c_2}{81}P_s^8&\ge&0,\\
\label{cond2}
b_1+\frac{c_1}{3}P_s^2+\frac{2}{3}b_2P_s^4+\frac{2c_2}{81}P_s^8&\le&0.
\end{eqnarray}

The parametric equations for the boundaries of the phase stability
domain are obtained by replacing the inequality in either
(\ref{cond1}) or (\ref{cond2}) with an equality and solving it
together with~(\ref{st}). The line resulting from (\ref{st}) and
(\ref{cond1})

\begin{equation} \label{rho} \left\{
\begin{array}{l}
\displaystyle
a_1(P_s)=\frac{c_1}{9}P_s^4+\frac{8b_2}{9}P_s^6+\frac{8c_2}{243}P_s^{10},\\
\displaystyle
b_1(P_s)=-3a_2-\frac{c_1}{3}P_s^2-2b_2P_s^4-\frac{5c_2}{81}P_s^8.
\end{array}\right.
\end{equation}
has a cusp at $c_1<0$ shown in the top inset of
Fig.~\ref{fig1}(a). This feature is defined by the parametric
equation

\be \label{cusp} c_1+12b_2P_{cusp}^2+\frac{20c_2}{27}P_{cusp}^6=0.
\ee
 This equation has no real solutions for $c_1>0$, as the cusp
merges with the $a_1$ axis. Another interesting feature of the
phase diagram is a first-order transition line below the $b_1$
axis, shown in the top inset of Fig.~\ref{fig1}(a). It corresponds
to an \textit{isostructural} transition between two phases with
the same structure and symmetry R, with the value of $P_s$ being
the only difference between them~\cite{larin}. Let $P_s'$
and $P_s''$ be the two values obtained from the equation of
state~(\ref{st}). Then the transition line is defined by
$F_R(P_s')=F_R(P_s'')$, where $F_R(P_s)$ is the potential of phase
R:
\begin{eqnarray}
\label{eqr}
F_R(P_s)&=&a_1P_s^2+\frac{(3a_2+b_1)}{3}P_s^4+\frac{c_1}{27}P_s^6+\frac{b_2}{9}P_s^8\nonumber\\
&& +\frac{c_2}{729}P_s^{12}.
\end{eqnarray}

All the equations relating to isostructural phases have an
additional permutation symmetry ${P_s'}^2\leftrightarrow
{P_s''}^2$. Introducing new variables $U=({P_s'}^2+{P_s''}^2)/3$
and $V={P_s'}^2{P_s''}^2/9$, which are invariant under the
permutation, we obtain

\begin{widetext}
\begin{equation}\label{istr}
\left\{
\begin{array}{l}
6a_1-c_1(U^2+2V)-18b_2U^3-2c_2U^3(2U^2-5V)=0,\\
a_1-4(3a_2+b_1)U-c_1(2U^2+V)-12b_2U(2U^2-V)-2c_2U^3(2U^2-5V)=0,\\
2(3a_2+b_1)+c_1U+12b_2(U^2-V)+2c_2(U^4-3U^2V+V^2)=0.
\end{array}\right.
\end{equation}
\end{widetext}
$V$ can be eliminated from~(\ref{istr}) and resulting equations
can be solved for $a_1$ and $b_1$. One of the solutions
parametrically defines an isostructural R--R transition line by

\begin{equation} \label{isostr} \left\{
\begin{array}{l}
\displaystyle
a_1(U)=\frac{c_1^2}{18c_2U}+\frac{b_2c_1}{c_2}+\frac{c_1}{9}U^2-{2b_2U^3}
-\frac{4c_2}{9}U^5,\\
\displaystyle
b_1(U)=-\frac{c_1^2}{36c_2U^2}-3a_2+\frac{9b_2^2}{c_2}-\frac{2c_1}{9}U\\
\displaystyle \qquad\qquad +3b_2U^2+\frac{5c_2}{9}U^4.
\end{array}\right.
\end{equation}
A similar isostructural transition line appears in phase M$_A$ and
is shown enlarged in the bottom inset of Fig.~\ref{fig1}.

The first-order R--C transition occurs when the two phases have
equal potentials: $F_R(P_s)=0$. This condition together with
equation of state~(\ref{st}) gives
\begin{equation} \label{rhotr} \left\{
\begin{array}{l}
\displaystyle
a_1(P_s)=\frac{c_1}{27}P_s^4+\frac{2b_2}{9}P_s^6+\frac{4c_2}{729}P_s^{10},\\
\displaystyle
b_1(P_s)=-3a_2-\frac{2c_1}{9}P_s^2-b_2P_s^4-\frac{5c_2}{243}P_s^8.
\end{array}\right.
\end{equation}

We would like to point out some other important features of the
phase diagram of Fig.~\ref{fig1}, which can be represented in a
simple analytical form. For $c_1>0$, phase M$_C$ lies between
T--M$_C$ and O--M$_C$ second-order phase boundaries defined
respectively by

\begin{equation} b_1=0, \text{ and } b_1=-\frac{a_1^2b_2}{8a_2^2}.
\end{equation}
The second-order O--M$_B$ transition line is parabolic in the
$(a_1, b_1)$ parameter space

\be (2b_2a_1+c_1b_1)^2-(c_1^2+8a_2b_2)(c_1a_1-4a_2b_1)=0. \ee

The lowest symmetry Tri phase appears at $c_1<0$ between phases
M$_A$ and M$_B$ in a slice restricted by a parametrically defined
line

\begin{equation} \label{tr} \left\{
\begin{array}{l}
\displaystyle a_1(t)=-4a_2t^2+\frac{c_1a_2}{c_2t^4},\\
\displaystyle b_1(t)=-2b_2t^4+\frac{2c_1b_2}{c_2t^2},
\end{array}\right.
\end{equation}
where $t$ is a real parameter. The coordinates of the four-phase
R--M$_B$--M$_A$--Tri critical point can be obtained
from~(\ref{tr}):

\begin{equation} \label{Fpoint} a_1=3a_2\left(\frac{4c_1}{c_2}\right)^{1/3};
\qquad b_1=-3b_2\left(\frac{2c_1^2}{c_2^2}\right)^{1/3}.
\end{equation}

The first-order transition lines M$_B$--R, R--M$_A$, M$_A$--T, and
R--O, and isostructural M$_A$--M$_A$ line were obtained
numerically, as the method described in
Eqs.~(\ref{st})--(\ref{rhotr}) in these cases leads to high order
equations for $a_1$ and $b_1$.

All possible second-order phase transitions are shown in
Fig.~\ref{fig1}(c). In agreement with the general
analysis~\cite{sakh}, only phases R and T are accessible from
phase C via a second-order transition. The M$_C$--Tri second order
transition occurs in the $c_1=0$ plane and is not shown in
Fig.~\ref{fig1}.

In Sec.~\ref{sec3} we will give special attention to some other
features of the phase diagram shown in the lower inset of
Fig.~\ref{fig1}(a). They are related to the structure of the phase
diagrams, recently observed in PZT~\cite{noh3,noh4} and
PMN-PT~\cite{ye}.

\section{Phase diagrams of PZT, PMN-PT and PZN-PT}
\label{sec3}

The seemingly simple twelfth-order model potential $F_{init}$ from
Eq.~(\ref{prim}) has led, as can be seen from Fig.~\ref{fig1}, to
a complex phase diagram describing all eight phases allowed by the
symmetry of the problem. All possible second-order phase
transitions among the proper ferroelectric phases in perovskites
are established by Fig.~\ref{fig1}(c), so any approximations made
to describe transitions driven by the ferroelectric order
parameter should not add or forbid any second-order transition
lines obtained in the model~(\ref{prim}). On the other hand,
first-order phase transitions between any two of the eight phases
can be expected, since, in general, there are no arguments that
can forbid these transitions. For example, a direct O--T
transition has been experimentally observed in PZN-PT~\cite{noh5}
and earlier in BaTiO$_3$.

\begin{figure*}
\includegraphics{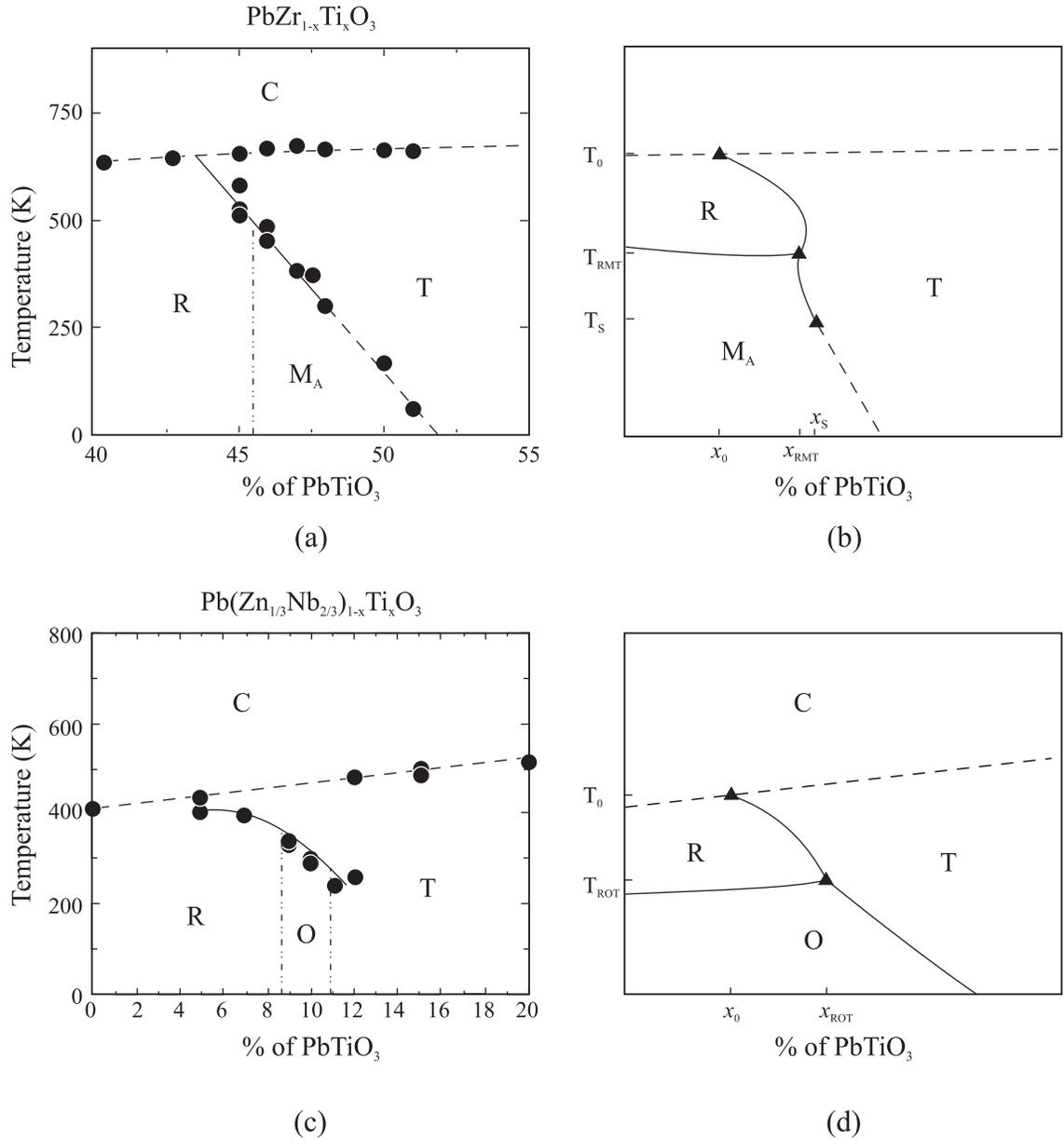}
\caption{\label{fig2}(a), (c) Phase diagrams of PZT and PZN-PT
from Refs.~\onlinecite{noh3,noh4,jaffe,kuwata}. (b), (d) Phase
diagrams calculated based on the model potentials $F_{PZT}$ and
$F_{PZN}$, as described in the text. Critical points are shown as
triangles, their coordinates are marked on axes.}
\end{figure*}

In Fig.~\ref{fig2} we plot the experimental phase diagrams of PZT
and PZN-PT together with our calculations based on the model
described. In Sec.~\ref{sec2} we have shown that the triple point
of phases R, T, and M$_A$ can appear as an intersection of the
first-order phase transition lines (see Fig.~\ref{fig1}(a)). A
second-order phase transition M$_A$--T is possible below the
critical point $S$. This is exactly the topological structure of
the M$_A$--T phase transition line that has been observed in
PZT~\cite{noh3}. To obtain reasonable agreement with the
experiment, we assumed that the coefficients $a_1$ and $b_1$
in~(\ref{prim}) are linear functions of temperature and
composition, while the rest of the parameters are constant:

\be
\begin{array}{rcl}
a_1&=&\alpha_T(T-T_0)+\alpha_x(x-x_0),\\
b_1&=&\beta_T(T-T_0)+\beta_x(x-x_0).
\end{array}
\ee

Keeping in mind the global topology of the phase diagram shown in
Fig.~\ref{fig1}, we can identify the features related to the
piezoelectric compounds. When applying the developed model to the
particular systems, we add some symmetry allowed terms to the
expansion~(\ref{prim}). This enables us to use simple temperature
and concentration dependencies of the phenomenological parameters
while achieving a good fit to the experimental phase diagrams. To
model a PZT-type phase diagram (also related to PMN-PT), we use an
expansion for the Landau potential

\be \label{pot1} F_{PZT}=F_{init}+d_{12}J_1J_2. \ee

Upon this modification, the M$_A$--T second-order transition line
remains straight but becomes tilted with respect to the $a_1$
axis:

\be b_1=\frac{d_{12}a_1}{2a_2}. \ee

Since the phenomenological phase diagram can always be
isomorphically transformed to fit the experimental data, we do not
try to obtain a perfect quantitative fit. Rather, we calculate the
coefficients in arbitrary units to reproduce the qualitative
features of the phase diagrams. The calculated values are shown in
Table~\ref{tbl}.

\begin{table}
\caption{\label{tbl}Parameters of the Landau potentials used for
qualitative fit of phase diagrams for PZT and PZN-$x$PT.}
\begin{ruledtabular}
\begin{tabular}{ccc}
  Parameter & Value for $F_{PZT}$ & Value for $F_{PZN}$\\
\hline
  $\alpha_T$ & 1 & 1\\
  $\alpha_x$ & -0.07 & -0.8\\
  $\beta_T$ & 0.05 & 0.6\\
  $\beta_x$ & 1 & 1\\
\hline
  $c_1$ & -25 & -40\\
  $a_2$ & 9 & 9 \\
  $b_2$ & 0.8 & 0.7 \\
  $c_2$ & 3.6 & 0\\
  $d_{12}$ & -0.3 & 8 \\
  $d_{13}$ & 0 & 13 \\
\end{tabular}
\end{ruledtabular}
\end{table}

We used a relatively large absolute value for $c_1$ to obtain
comparable lengths of the first-order M$_A$--T transition line and
R--T phase boundary, modelling the experimental data for PZT. This
results in a vanishingly small region of coexistence of the two
isostructural M$_A$ phases (see the bottom inset in
Fig.~\ref{fig1}(a)). The critical point $S$ then almost merges
with point $L$, the latter being a tangency point of the T--M$_A$
second-order transition line continuation and the stability
boundary of phase M$_A$.

For the expansion~(\ref{pot1}), point $L$ has coordinates:

\be a_1=\frac{2a_2^2c_1}{4a_2b_2-d_{12}^2}; \quad
b_1=\frac{a_2c_1d_{12}}{4a_2b_2-d_{12}^2}. \ee

More extensive modifications of $F_{init}$ have been used to
accommodate the features of the experimental phase diagram of
PZN-PT, namely a direct O--T transition and, more importantly, the
separation of the triple points R--O--T and C--R--T:

\be \label{pot2} F_{PZN}=F_{init}+d_{12}J_1J_2+d_{13}J_1J_3. \ee

We find that the part of the global phase diagram, relevant to
PZN-PT, is almost unaffected by the variation of $c_2$. Therefore,
we assume $c_2=0$. The other coefficients are found in
Table~\ref{tbl}.

\section{Piezoelectric properties}
\label{sec4}

Elastic energy and electrostrictive coupling have to be introduced
into Landau potential to obtain the components of the
piezoelectric tensor $\mathsf{d}_{ik}$. The piezoelectric
constants are inversely proportional to combinations of the second
derivatives of $F_{PZT}$ and $F_{PZN}$ with respect to $P_x, P_y,
P_z$~\cite{cross}. Some of these combinations vanish at the
stability boundaries of the phases, resulting in anomalous values
of the piezoelectric constants.

By analyzing the phase stability conditions obtained in
Sec.~\ref{sec2}, we find the components $\mathsf{d}_{ik}$ which
can be large for each phase of the phase diagrams Fig.~\ref{fig2}.

In the R phase, all four independent piezoelectric constants are
expected to be large at all the phase transition lines with phases
T, M$_A$, and O.

The tetragonal piezoelectric $\mathsf{d}_{15}$ (in this section we
use a pseudo-cubic orthogonal basis in the $\mathsf{d}_{ik}$
notation) component of the piezoelectric tensor can be large at
the T--R, T--M$_A$ and T--O phase boundaries, while
$\mathsf{d}_{33}$ and $\mathsf{d}_{31}$ are not expected to have
anomalies in temperature and composition dependence at these
lines.

A complex behavior of the piezoelectric constants is expected in
phase M$_A$ at the boundary with phase T. In the second-order
transition part of this phase boundary, no anomalies are expected.
This is because no stability condition is violated there, even
though it is a bifurcation line where the monoclinic solution of
the equations of state becomes imaginary. However, this situation
changes drastically when M$_A$--T is a first-order phase
transition. In this part of the M$_A$--T boundary and at the
M$_A$--R boundary, all ten monoclinic piezoelectric coefficients
can be large.

Another verifiable prediction of our theory is the connection
between the proximity of phase M$_A$ to the triclinic distortion
and anomalies in the monoclinic piezoelectric coefficients
$\mathsf{d}_{11}$, $\mathsf{d}_{12}$, $\mathsf{d}_{14}$, and
$\mathsf{d}_{15}$, at low temperatures.

Finally, $\mathsf{d}_{33}$ and $\mathsf{d}_{31}$ can be large in
phase O at the T boundary, and $\mathsf{d}_{24}$ -- at the R
boundary.

\section{\label{sec5}Summary and discussions}

We have shown that the increasingly high-order phenomenological
models, used to describe the highly nonlinear piezoelectric
systems PZT, PMN-PT, and PZN-PT, are approximations of a complete
twelfth-order model~(\ref{full12}). Such a complex model is
difficult to analyze, but even the significantly simplified model
potentials $F_{PZT}$ and $F_{PZN}$ describe the phase diagram of
the compounds reasonably well, as seen in Fig.~\ref{fig2}. We have
allowed for some discrepancy between the calculated and the
experimental phase diagrams, for example, in the location of
R--M$_A$ and R--O phase boundaries, in order to keep the
dependence of the phenomenological parameters on temperature and
concentration as simple as possible. On the other hand, it is
clear, that since our phenomenological models correctly describe
the qualitative features of the phase diagram, a complete
quantitative agreement can be achieved by a careful fitting of the
phenomenological parameters to the experimental data.

The important features that
our simple models correctly reproduce are the separation of the triple
points R--O--T and R--M$_A$--T from the C--R--T triple point, and
the switching of the M$_A$--T phase transition between the first-
and the second-order types.

We also predict an anomalous behavior of certain piezoelectric
constants along the phase transition lines. These predictions
elucidate the nature of the ultra-high piezoelectricity and make
our theory easily verifiable. To our knowledge, systematic
measurements of evolution of the piezoelectric constants in the
vicinity of the newly found phase boundaries in PZT, PMN-PT, and
PZN-PT have not been done yet.

Finally, group-theoretical and catastrophe-theory methods of phase
transition theory described in this paper stand on their own. We
described a general approach that puts phenomenological analysis
on firm footing, and minimizes the necessary amount of numerical
calculations.

\begin{acknowledgments}
We would like to thank M. F. Kurpriyanov, E. S. Larin, D.
Vanderbilt, and G. Gaeta for very helpful comments and
discussions.
\end{acknowledgments}

\appendix\section{\label{apa}Calculation of IRBI for representations of space groups}

Let $\mathcal{L}$ be the group of all different matrices of a
$p$-dimensional representation of a space group. The
$\mathcal{L}$-group can be thought of as a point group in the space
of the order parameter components $\eta_i, i=1,\ldots, p$. The Landau
potential is invariant under transformations of this group.

\emph{Integrity rational basis of invariants}(IRBI) of the
$\mathcal{L}$-group is a minimal set of homogeneous
$\mathcal{L}$-invariant polynomials depending on $\eta_i$, such
that any other invariant homogeneous polynomial is an algebraic
combination of the elements of this set. One of the methods to find
IRBI is described briefly below~\cite{book}.

It can be proved that for every group $\mathcal{L}$ the
following sequence of subgroups can be constructed:

\begin{equation}
\label{normal}
E\equiv\mathcal{L}_0\subset\mathcal{L}_1\subset\ldots\subset\mathcal{L}_N\equiv\mathcal{L},
\end{equation}
where each $\mathcal{L}_i$ is an invariant subgroup of
$\mathcal{L}_{i+1}$ and $E$ -- identity group.

The problem of constructing IRBI of $\mathcal{L}$ is reduced to
constructing invariants of the factor groups
\begin{equation}\label{quot}
  A_i=\mathcal{L}_i/\mathcal{L}_{i-1}
\end{equation}
at every step of~(\ref{normal}). It turns out that for any
$\mathcal{L}$-group, $\mathcal{A}_i$ acts on the invariants
obtained in the previous $(i-1)$-th step as cyclic group of order
2 or 3, for which the invariants can be easily constructed.
Removing the algebraic dependencies among thus obtained
polynomials at each step, at $N$-th step one obtains the IRBI for
the given $\mathcal{L}$-group.

If $m$ is the number of invariants, constituting the IRBI for a
space group representation then, in general, $m\ge p$, and there
exist $m-p$ algebraic relations of higher than first order among
the $m$ invariant polynomials. However, for irreducible
representations of groups generated by reflections $m=p$. This is
the situation for the ferroelectric order parameter in $Pm3m$
group, for which the step-by-step calculation of invariants
results in the IRBI~(\ref{irbi}).

The method described here allows for a simple generalization to
groups with continuous subgroups, such as a gauge transformation
group~\cite{book,supc}.

\section{\label{apb}Structural stability of phenomenological potentials}

In this appendix we review the method based on catastrophe theory,
which is used to analyze the stability of the potential functions.
We consider the simplest case of an irreducible representation of
a group generated by reflections, for which $m=p$ (see
Appendix~\ref{apa}).

To determine the structural stability of some potential function $F(J_1,\ldots,J_m)$, we
need to introduce the algebraic combinations:
\begin{equation} \label{combin}
U_i(J_1,\ldots,J_m)=\sum_{k=1}^m\frac{\partial F}{\partial
J_k}(\nabla J_k, \nabla J_i), i=1,\ldots,m,
\end{equation}
where $\nabla J_i$ is the gradient of invariant $J_i$ in the order
parameter space. Scalar products that appear in~(\ref{combin}) can
always be expressed in terms of invariant polynomials
$J_1,\ldots,J_m$. A term can be omitted in the potential function
without violating the type of the extremal behavior of this
function if its coefficient is small and if it can be expressed as

\begin{equation}
\label{term}
\sum_{i=1}^m\mathcal{P}_i(J_1,\ldots,J_m)U_i(J_1,\ldots,J_m)+\text{
h. o. t.}
\end{equation}
Here $\mathcal{P}_i(J_1,\ldots,J_m)$ are some polynomials and  `h.
o. t.' stands for `higher order terms'.

Following this algorithm, we obtain for the integrity
basis~(\ref{irbi}) and potential~(\ref{prim})

\begin{equation}
\begin{array}{ll}
(\nabla J_1)^2=4J_1, & (\nabla J_1,\nabla J_2)=8J_2,\\
(\nabla J_2)^2=4J_1J_2+12J_3, & (\nabla J_2,\nabla J_3)=8J_1J_3,\\
(\nabla J_3)^2=4J_2J_3, & (\nabla J_1,\nabla J_3)=12J_3.
\end{array}
\end{equation}
and
\begin{eqnarray}
U_1&=&4a_1J_1+\text{ h. o. t.}\nonumber\\
U_2&=&8a_1J_2+\text{ h. o. t.}\nonumber\\
U_3&=&12a_1J_3+\text{ h. o. t.}.
\end{eqnarray}

It is now clear that every term in~(\ref{full12}), additional to
~(\ref{prim}), can be represented in the form~(\ref{term}).
However, there is another restriction that arises from the
requirement for a model to produce all the phases allowed by the symmetry
of the order parameter. As can be seen from Eq.~(\ref{tr}), triclinic
phase cannot appear in the phase diagram of the truncated below
the twelfth-order Landau potential expansion~(\ref{prim}). Consequently,
the potential~(\ref{prim}) is the simplest model that meets all the
requirements mentioned above.

\end{document}